\pdfoutput=1

\documentclass[11pt]{article}

\usepackage[]{acl}

\usepackage{times}
\usepackage{latexsym}

\usepackage[T1]{fontenc}

\usepackage[utf8]{inputenc}

\usepackage{microtype}

\usepackage{inconsolata}

\definecolor{green_colorblind_friendly}{HTML}{1b9e77}
\definecolor{orange_colorblind_friendly}{HTML}{d95f02}
\definecolor{purple_colorblind_friendly}{HTML}{7570b3}

\usepackage{booktabs, multirow, makecell}
\usepackage{enumitem}
\usepackage{xcolor,colortbl}
\usepackage{amsmath}
\usepackage{graphicx}
\usepackage{pifont}
\newcommand{\cmark}{\ding{51}}%
\newcommand{\xmark}{\ding{55}}%

\title{\mbox{\textsc{SyllabusQA}}: A Course Logistics Question Answering Dataset}

\author{Nigel Fernandez, Alexander Scarlatos, Andrew Lan  \\
  University of Massachusetts Amherst \\
  \texttt{\{nigel,ajscarlatos,andrewlan\}@cs.umass.edu}
}

\begin{document}
\maketitle

\begin{abstract}
Automated teaching assistants and chatbots have significant potential to reduce the workload of human instructors, especially for logistics-related question answering, which is important to students yet repetitive for instructors. However, due to privacy concerns, there is a lack of publicly available datasets. We introduce \mbox{\textsc{SyllabusQA}}~\footnote{Dataset and code: \url{https://github.com/umass-ml4ed/SyllabusQA}}, an open-source dataset with $63$ real course syllabi covering $36$ majors, containing $5,078$ open-ended course logistics-related question-answer pairs that are diverse in both question types and answer formats. Since many logistics-related questions contain critical information like the date of an exam, it is important to evaluate the factuality of answers. We benchmark several strong baselines on this task, from large language model prompting to retrieval-augmented generation. We introduce Fact-QA, an LLM-based (GPT-4) evaluation metric to evaluate the factuality of predicted answers. We find that despite performing close to humans on traditional metrics of textual similarity, there remains a significant gap between automated approaches and humans in terms of fact precision.
\end{abstract}


\section{Introduction}

In educational applications, artificial intelligence (AI) approaches have shown significant promise in improving learning outcomes \cite{aleven2016instruction,vanlehn2011relative}, by automatically providing feedback to students or engaging in tutoring dialogues with them. The key idea is to use AI to create an on-demand virtual \emph{teaching assistant} to interact with many students simultaneously; see, e.g., Khamigo from Khan Academy \cite{khamigo}. These approaches can scale up the effort of expert human teachers and tutors, and relieve them from doing repetitive tasks so that they can focus on providing personalized feedback or designing new learning content \cite{bazaar}. In higher education, one promising avenue for AI-powered teaching assistants to reduce human effort is \emph{course logistics-related} question answering (QA): answering student questions on logistics whose answers can be directly found or inferred from the syllabus.

\begin{table}
\small
\centering
\begin{tabular}{p{0.95\linewidth}}

\toprule

\textbf{Course Name:} BIOL 151: Introductory Biology\\
\textbf{Length:} $5864$ tokens, $16$ pages\\
\textbf{Formatting:} Tables, lists, schedules, images\\
\textbf{Syllabus Text:}\\
\textcolor{purple_colorblind_friendly}{[page 5]} \ldots The two mid-term exams are not cumulative 
\ldots The final exam is cumulative \ldots\\
Exam Replacement: If you score higher on the final exam than on one of your mid-term exams, the final exam score will replace that mid-term exam score. This is only true if you take both.
\\

\ldots\\
\textcolor{purple_colorblind_friendly}{[page 9, 10]} \ldots Topics that will be covered in the order shown below and we will spend roughly a week on each topic. 1. The cell \ldots\ Detailed topic list 1) The Cell a) Organization b) membranes \ldots\\

\midrule
\textcolor{purple_colorblind_friendly}{[multi-hop reasoning][implicit][page 5]}\\
\textcolor{orange_colorblind_friendly}{Q}: If I don't do very well on one of my midterms, will I be able to pull up my grade somehow?\\
\textcolor{green_colorblind_friendly}{A}: If you get a higher grade on your final exam than on your midterm, the midterm grade will be dropped giving you a higher overall grade for the course.\\

\midrule
\textcolor{purple_colorblind_friendly}{[single-hop reasoning][implicit][page 9, page 10]}\\
\textcolor{orange_colorblind_friendly}{Q}: If I have to miss the first few days of the class, what content will I miss?\\
\textcolor{green_colorblind_friendly}{A}: If you miss the first few days of class, you will miss learning about the cell, which includes organization and membranes.\\ 


\midrule
\textcolor{purple_colorblind_friendly}{[adversarial]}\\
\textcolor{orange_colorblind_friendly}{Q}: Are there any prerequisites for this course?\\
\textcolor{green_colorblind_friendly}{A}: No/insufficient information\\

\bottomrule

\end{tabular}
\caption{Syllabus and Question-Answer examples from \mbox{\textsc{SyllabusQA}}. Each QA pair has meta info (answer spans or reasoning steps) and may span multiple pages.}
\label{tab:syllabus_qa_example}
\end{table}

\begin{table*}[htbp]
\small
\centering
\scalebox{.78}{
\begin{tabular}{p{0.31\linewidth}p{0.02\linewidth}p{0.03\linewidth}p{0.075\linewidth}p{0.03\linewidth}p{0.06\linewidth}p{0.05\linewidth}p{0.1\linewidth}p{0.05\linewidth}p{0.125\linewidth}p{0.13\linewidth}} 
\toprule

Dataset & Edu. & Public & Domain Diversity & Long Doc. & Complex Format & Adv.\ Q & Q Type & A Type & Generation & Data Source\\

\midrule

\textsc{ChaTa} \cite{hicke2023chata} & \cmark & \xmark & \xmark & \xmark & \xmark & \xmark & Open-ended & Natural & Student-written & Piazza\\

\textsc{FairytaleQA} \cite{xu-etal-2022-fantastic} & \cmark & \cmark & \cmark & \xmark & \xmark & \xmark & Open-ended & Natural & Expert & Literature\\

\textsc{Book Test} \cite{childrens-book-test} & \xmark & \cmark & \cmark & \xmark & \xmark & \xmark & Cloze & Span & Automated & Literature\\

\textsc{NarrativeQA} \cite{narrativeqa} & \xmark & \cmark & \cmark & \cmark & \xmark & \xmark & Open-ended & Natural & Crowd-sourced & Movies/literature\\

\textsc{CliCR} \cite{clicr} & \xmark & \cmark & \xmark & \cmark & \xmark & \xmark & Cloze & Span & Automated & Medical reports\\

\textsc{NewsQA} \cite{newsqa} & \xmark & \cmark & \xmark & \xmark & \xmark & \cmark & Open-ended & Span & Crowd-sourced & News\\

\textsc{QuAC} \cite{quac} & \xmark & \cmark & \cmark & \xmark & \xmark & \cmark & Open-ended & Span & Crowd-sourced & Wikipedia\\

\textsc{OpenBookQA} \cite{openbookqa} & \xmark & \cmark & \cmark & \cmark & \cmark & \xmark & MCQ & Multi & Crowd-sourced & Science books\\

\textsc{HotpotQA} \cite{hotpotqa} & \xmark & \cmark & \cmark & \cmark & \xmark & \cmark & Open-ended & Span & Crowd-sourced & Wikipedia\\

\midrule

\mbox{\textsc{SyllabusQA}} & \cmark & \cmark & \cmark & \cmark & \cmark & \cmark & Open-ended & Natural & Crowd-sourced & Course syllabi\\

\bottomrule
\end{tabular}
}
\caption{Comparison of \mbox{\textsc{SyllabusQA}} with existing educational or generic QA datasets.}
\label{tab:dataset_comparison}
\end{table*}

There exist many approaches for automated QA in online courses (both logistics-related and content-related), using tools from rule-based AI systems \cite{ryan,feng2006intelligent} and expert systems with knowledge bases \cite{goel2017using,goel2021agent} to end-to-end text generation \cite{zylich2020exploring}. Recently, large language model (LLM)-based approaches have shown great promise to improve the coverage and answer quality over traditional QA approaches \cite{hicke2023chata}. See Section~\ref{sec:appendix_rw} in the Supplementary Material for a detailed discussion on related work. Unfortunately, these approaches are mostly developed and evaluated on proprietary data due to student privacy concerns, which prevents more researchers from contributing to the development of automated QA systems for education.

\begin{figure*}[htbp]
\centering
\includegraphics[width=1\linewidth]{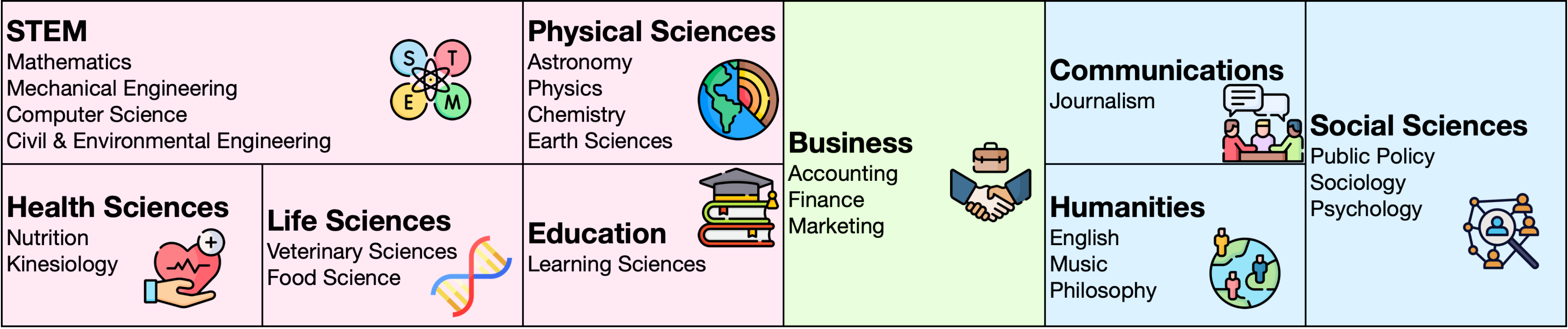}
\caption{Domain diversity in \mbox{\textsc{SyllabusQA}} covering $36$ majors. For visual clarity, we show representative majors.} 
\label{fig:diverse_majors}
\end{figure*}

For evaluation, especially in logistics-related QA that often contains critical information, the factuality of predicted answers is more important than measuring surface textual features. Moreover, text similarity metrics may not be suitable for some open-ended natural language generation tasks~\cite{amidei-etal-2018-rethinking}. As an example, the answer ``The final exam will be on Dec 15'', has high surface-level textual similarity with the reference answer, ``The final exam is on Dec 14'', but contains a critical factual error that may lead to significant negative consequences to students. 
Meanwhile, human instructors and teaching assistants often answer student questions in a concise way, without giving any unnecessary information. Therefore, it is important for LLM-based approaches to generate answers that are both concise and precise.

\paragraph{Contributions}
In this paper, we introduce the \mbox{\textsc{SyllabusQA}} dataset for course logistics-related QA. We publicly release this dataset and hope that it can be a benchmark for future work on developing and evaluating automated QA approaches for teaching assistance. Our contributions are:

\noindent\textbf{First}, we collect a diverse set of $63$ real course syllabi covering $36$ majors across $12$ universities, and employ crowd annotators to write $5,078$ logistics-related QA pairs with the goal of simulating what students would ask in a real-world course. 

\noindent\textbf{Second}, we detail the diverse composition of syllabi and QA pairs in \mbox{\textsc{SyllabusQA}}, in terms of syllabi domain, question types, answer sources, and different language styles. We lay out metrics to evaluate different aspects of open-ended automated QA approaches, in terms of both surface textual similarity and more importantly, the factuality of predicted answers grounded in the syllabus. 

\noindent\textbf{Third}, we conduct extensive experiments to benchmark the QA performance of several strong baselines on \mbox{\textsc{SyllabusQA}}. Overall, LLM-based approaches perform similar to humans on surface-level textual similarity metrics but worse on factuality metrics. We found that fine-tuning on real QA pairs from \mbox{\textsc{SyllabusQA}} performs much better than LLM prompting approaches and that retrieval-augmented generation is especially important.

To the best of our knowledge, \mbox{\textsc{SyllabusQA}} is the first publicly available real-world course logistics-related QA dataset. \mbox{\textsc{SyllabusQA}} assesses automated QA models on various natural language understanding aspects including handling challenging question types, from reasoning-based ones to adversarial ones, understanding of long input documents sourced from diverse course majors, processing complex input formatting including tables and schedules, and answering open-ended questions in a similar way as human instructors and TAs. Table~\ref{tab:dataset_comparison} highlights the difference between \mbox{\textsc{SyllabusQA}} and existing datasets; see also Section~\ref{sec:comparison} for a detailed discussion.


\section{\mbox{\textsc{SyllabusQA}} Data Collection}

We now detail how we construct \mbox{\textsc{SyllabusQA}} to address some of the limitations in existing datasets.


\subsection{Source Syllabi}

The source texts we include in this dataset are anonymized course syllabi, based on which QA pairs are generated. We collected $63$ unique syllabi from instructors across $12$ unique universities worldwide, at both undergraduate and graduate levels. 
We include a wide range of diverse course subjects, from science and engineering to humanities and business, as shown in Figure~\ref{fig:diverse_majors}. We also collected syllabi from courses with different formats: lab-based ones, project-based ones, lecture-based ones, etc. 
To ensure anonymity, we manually replaced personally identifiable information on instructors and teaching assistants with placeholders, along with any private information like platform login codes. On average, there are $8$ pages per syllabus, with a minimum of $2$ and a maximum of $27$. We store all syllabi in their raw PDF format. Apart from the considerable variation in length, the diverse information formats in the syllabi, such as lists, tables, boxes, schedules, forms, and even diagrams and images, make QA on \mbox{\textsc{SyllabusQA}} especially challenging: doing so requires techniques ranging from parsing, information retrieval, to long-document QA.


\subsection{Design Considerations}
\label{subsection:q_type}

\paragraph{Categorization by Question Types}
\textsc{SyllabusQA} is designed to include diverse question types across common logistics-related questions that human instructors and teaching assistants often face. There are $7$ question types defined below, with real examples of each type:
\begin{itemize}[noitemsep,topsep=0.2pt]
    \item \textbf{Yes/No}: questions that have answers consisting of a single word {yes/no} answer, e.g., ``Is there a separate lab section for this class?''
    \item \textbf{Single-factual}: questions that have answers containing a single explicit fact from the syllabus, e.g., ``When are office hours?''
    \item \textbf{Multi-factual} questions that have answers combining multiple explicit facts from the syllabus, e.g., ``What software and applications are used in this class?''
    \item \textbf{Single-hop reasoning}: questions that have implicit answers requiring a single reasoning step, e.g., ``I have not yet taken Advanced Biology, can I still take this course?''
    \item \textbf{Multi-hop reasoning}: questions that have implicit answers requiring multiple reasoning steps, e.g., ``Can I start the class six weeks in and get an A?''
    \item \textbf{Summarization}: questions that have answers that require summarizing information from multiple parts of the syllabus, e.g., ``Could you tell me how class participation grades are broken down?''
    \item \textbf{Adversarial}: questions that have no answers due to insufficient information, e.g., ``Can I contact the instructor over Zoom?''
\end{itemize}

\paragraph{Categorization by Answer Source}
Questions in \mbox{\textsc{SyllabusQA}} can also be categorized by the source of their answers as defined below:
\begin{itemize}[noitemsep,topsep=0.2pt]
    \item \textbf{Explicit}: questions that have answers directly from the syllabus, including question types Yes/No, Single-factual, and Multi-factual.
    \item \textbf{Implicit}: questions that have answers not explicitly in the syllabus and require inference, including question types Single-hop reasoning, Multi-hop reasoning, and Summarization.
    \item \textbf{Insufficient Information} questions that do not have answers due to insufficient supporting information in the syllabus, including questions of type Adversarial.
\end{itemize}

The combination of multiple question types and answer sources makes \mbox{\textsc{SyllabusQA}} challenging for automated QA approaches. The involvement of adversarial questions, in particular, requires models to not hallucinate information not present in the syllabus, which is a critical test to pass before any potential deployment to real students.


\subsection{Annotation Procedure}
\label{sec:annotationprocedure}

\paragraph{Annotation Guidelines}
For simplicity, we use a straightforward annotation task: given a syllabus, we ask an annotator to simulate logistics-related QA pairs seen in real course classrooms. They write $2$ QA pairs for each of the $7$ question types for a total of $14$ QA pairs, to ensure QA pairs are evenly distributed across question types. Before the task, we guide them through a tutorial to help them get familiar with the task and write diverse QA pairs. We give them a list of $14$ diverse QA examples manually written by us, covering a wide spectrum of QA examples seen in classrooms, with $2$ examples for each question type, on a single example syllabus. We encourage annotators to write open-ended questions with answers written in their own words, giving them the flexibility to come up with questions that are difficult to answer. 
See Section~\ref{sec:appendix_dataset_collection} in the Supplementary Material for the instructions we provide. 

For explicit questions with answers directly found in the syllabus, including the Yes/No, Single-Factual, and Multi-Factual types, an annotator additionally provides one, one, and up to five answer spans respectively, i.e., snippets directly copied from the reference syllabus, supporting their answer.
For Single-hop and Multi-hop reasoning questions, an annotator additionally provides one and up to five reasoning steps respectively, written in their own words to detail their thought process before arriving at the final answer.
For Summarization questions, an annotator additionally provides up to five answer spans from the syllabus used to construct the summary. This additional information provided by annotators encourages their answers to be cohesive and faithful to the reference syllabus and can be further used to aid development of automated QA approaches. For Adversarial questions, annotators do not write answers and we use ``No/insufficient information'' as the answer.

\paragraph{Crowdworker Recruitment}
\mbox{\textsc{SyllabusQA}} is designed to include questions written in different styles. Therefore, we recruited a large pool of over $200$ annotators across two popular platforms, Amazon Mechanical Turk (AMT) and Prolific, all with an undergraduate bachelor's degree or above, located in the United States or Canada. 
To ensure diversity in language style, we asked each annotator to write a maximum of $14$ QA pairs per syllabus, and a maximum of $112$ total QA pairs across various syllabi. We collected this data over $5$ months.

\paragraph{Quality Control} To ensure high data quality, we asked each annotator on AMT to write $14$ QA pairs on an assigned syllabus in a screening session. We only include data from those who pass the screening test in our dataset and invite them for additional annotation rounds. For annotators on Prolific, in addition to Prolific's screening process, we manually checked a random subset of QA pairs written by each annotator and found that their data meets our quality standards. For further quality control, we also run heuristics-based checks to filter out unsuitable QA pairs.
Since our source syllabi are anonymized, we discard questions asking for personally identifiable information about instructors and/or teaching assistants. We also post-process answers to Yes/No type questions to a single word since some contain additional explanations.

\paragraph{Agreement among Annotators}
To aid development of automated QA approaches, we explicitly aimed at encouraging language diversity across QA pairs in \mbox{\textsc{SyllabusQA}} by employing a large pool of annotators to reflect the diversity in real questions asked by students in classrooms. Therefore, inter-annotator agreement may be low, especially if evaluated on traditional text similarity measures. To formally measure this agreement, we had an independent expert annotator with extensive teaching experience as a college instructor, write \emph{oracle} answers for around $20\%$ of randomly sampled questions from the test set. We provide this expert with the same instructions and tutorial as the annotators and ask them to first select the question type before writing the answer accordingly. 

Following related work on open-form QA datasets~\cite{xu-etal-2022-fantastic}, we do not use traditional inter-annotator agreement metrics since we asked annotators to write QA pairs in their own words and style, which makes these metrics unsuitable~\cite{amidei-etal-2018-rethinking}. Instead, we mainly evaluate agreement on \emph{factual information overlap}, which measures the overlap in key facts contained in the annotator's answer and an oracle answer. We use a GPT-4-based evaluation of precision and recall of factual information present between the answers written by the expert and annotators, which we detail later in Section~\ref{sec:metric}. This metric results in an inter-annotator precision of $0.707$ and recall of $0.664$, indicating high overlap in factual information. As a reference, agreement measured in traditional surface textual similarity metrics is $0.419$ in ROUGE-L F1~\cite{lin-2004-rouge} and $0.684$ in BERTScore~\cite{bert-score}. This result is not surprising since we found that although annotator-written answers have varied surface language styles, they contain very similar key information.

\paragraph{Qualitative Analysis of Ground-truth Answers}
We analyze annotator-written ground-truth answers where an independent expert annotator manually verified a random sample of QA pairs from $10\%$ of the test set on: 1) precision and 2) recall. For precision, we test if the ground-truth answer is relevant to the question and supported by the syllabus. For recall, we test if the ground-truth answer contains all the relevant information required to answer the question. We found that $82\%$ ($77\%$) of ground-truth answers have perfect precision (recall), $11\%$ ($16\%$) of ground-truth answers have partial precision (recall), and $7\%$ ($7\%$) of ground-truth answers have poor precision (recall). Through a qualitative error analysis, we find the following error types leading to less-than-perfect precision/recall in ground-truth answers: 1) imperfect annotator 
recall/human error, 2) ambiguous syllabus information, and 3) ambiguous annotator-written answers. We provide details and examples in Section~\ref{sec:appendix_analysis_ground_truth_answers} in the Supplementary Material.


\section{\textsc{SyllabusQA} Data Analysis}

\begin{table*}[htbp]
\small
\centering
\begin{tabular}{p{2.3cm}||c|c|c|c||c|c|c|c||c|c|c|c}

\toprule

\multirow{4}{*}{\mbox{\textsc{SyllabusQA}}}
& \multicolumn{4}{c||}{Train} & \multicolumn{4}{c||}{Validation} & \multicolumn{4}{c}{Test}\\
\cmidrule{2-13}
& \multicolumn{4}{c||}{$3018$ QA across $39$ Syllabi} & \multicolumn{4}{c||}{$957$ QA across $11$ Syllabi} & \multicolumn{4}{c}{$1103$ QA across $13$ Syllabi}\\
\cmidrule{2-13}
& $\mu$ & $\sigma$ & Min & Max & $\mu$ & $\sigma$ & Min & Max & $\mu$ & $\sigma$ & Min & Max\\

\midrule

\# pages / syllabus & $7.2$ & $4.6$ & $3$ & $25$ & $10.0$ & $8.6$ & $2$ & $27$ & $8.8$ & $6.2$ & $2$ & $23$\\
\# tokens / syllabus & $4.2$K & $2.5$K & $1.2$K & $12$K & $6.4$K & $6.2$K & $1.2$K & $20$K & $4.7$K & $3.1$K & $1.1$K & $10.7$K\\
\# tokens / question & $14.1$ & $6.8$ & $4$ & $83$ & $13.9$ & $6.3$ & $4$ & $50$ & $14.5$ & $6.6$ & $4$ & $55$\\
\# tokens / answer & $28.3$ & $30.3$ & $1$ & $281$ & $26.4$ & $28.8$ & $1$ & $203$ & $27.7$ & $29.6$ & $1$ & $246$\\

\bottomrule
\end{tabular}
\caption{Statistics of the \mbox{\textsc{SyllabusQA}} dataset which contains $5,078$ QA pairs grounded in $63$ syllabi.}
\label{tab:syllabus_qa_stats}
\end{table*}

\textbf{Statistics}
We randomly split the dataset into train-val-test splits with a proportion of roughly $60\%$-$20\%$-$20\%$ in terms of QA pairs. We ensure no overlap in syllabi across splits to test the generalization ability of QA approaches to answer questions based on unseen new syllabi. We show detailed statistics of \mbox{\textsc{SyllabusQA}} in Table~\ref{tab:syllabus_qa_stats}. Overall, there are $5,078$ QA pairs, almost uniformly distributed across $7$ question types. There are $2,177$ Explicit QA pairs, $2,181$ Implicit pairs, and $720$ Insufficient Information pairs. We also show a breakdown by question type in Section~\ref{sec:appendix_dataset_stats} in the Supplementary Material.

\paragraph{Question Diversity} 
We investigate question diversity in \mbox{\textsc{SyllabusQA}}, especially across questions based on the same reference syllabus. Questions with similar semantics across different syllabi are acceptable since their answers are grounded in different surrounding contexts in different locations. We compute the average intra-syllabus question pair similarity, then average across all syllabi, resulting in a ROUGE-L F1 score~\cite{lin-2004-rouge} of $0.126$, which indicates low intra-syllabus question similarity. This observation validates our design to encourage annotators to write diverse QA pairs with multiple question types and answer sources.

\subsection{Comparison to Existing Datasets}
\label{sec:comparison}

We compare \mbox{\textsc{SyllabusQA}} to existing datasets both in the education domain and for more general-purpose QA in Table~\ref{tab:dataset_comparison}. To the best of our knowledge, \mbox{\textsc{SyllabusQA}} is the first real publicly available course logistics-related QA dataset in educational settings; the data in \cite{labutov2018multi} is synthetically generated. ChaTA~\cite{hicke2023chata} uses a proprietary dataset (with a publicly available sample) based on QA pairs on Piazza from an introductory computer science course. In contrast, \mbox{\textsc{SyllabusQA}} is public and more diverse by design, covering $36$ majors. FairytaleQA~\cite{xu-etal-2022-fantastic} is a public dataset on reading comprehension QA from short fairytale story snippets of around $150$ tokens. In contrast, \mbox{\textsc{SyllabusQA}} covers logistics-related QA with much longer context since the syllabi average $8$ pages and $5$K tokens long. CliCR~\cite{clicr} tests QA on medical documents with automatically generated Cloze-type questions with document spans as answers. In contrast, SyllabusQA is crowdsourced with diverse types of open-ended QA pairs written simulating teaching assistants.

Compared to general-purpose QA datasets, NewsQA~\cite{newsqa} contains QA pairs grounded on news articles while QuAC~\cite{quac} has QA on multi-turn dialogues, both with answers as text spans. In contrast, SyllabusQA contains QA pairs grounded on long course syllabi, which require techniques from long document QA. It also contains questions with open-ended answers. NarrativeQA~\cite{narrativeqa} is sourced from movie scripts and books in plain text format. In contrast, SyllabusQA is sourced from course syllabi, containing different source text formats including tables and diagrams. HotpotQA~\cite{hotpotqa} focuses on multi-hop reasoning questions. In contrast, SyllabusQA has seven different question types including summarization, multi-factual, and adversarial. OpenBookQA~\cite{openbookqa} contains multiple-choice questions grounded on a set of facts. In contrast, SyllabusQA has different types of questions with open-ended answers.


\section{QA Performance on \mbox{\textsc{SyllabusQA}}}

We benchmark several strong baselines using state-of-the-art LLMs on \mbox{\textsc{SyllabusQA}}. For open-source LLMs, we use the popular LLaMA 2~\cite{touvron2023llama} model family, including the chat variants of LLaMA 2-7B, LLaMA 2-13B, and LLaMA 2-70B. Using open-source LLMs ensures the reproducibility of results and mitigates student privacy concerns in real-world deployment. For completeness, we also benchmark against GPT-4~\cite{openai2023gpt4} enabled with the retrieval assistant, a highly capable but proprietary LLM. We explore different approaches in our experiments including zero-shot prompting and supervised fine-tuning (SFT), with and without help from retrieval-augmented generation (RAG) techniques. We show all prompts in Section~\ref{sec:appendix_prompts} and model training details in Section~\ref{sec:appendix_params} in the Supplementary Material.

\subsection{Approaches}

In the basic \textbf{Zero-shot} approach, we prompt pre-trained LLaMA 2 models using only the question and system instructions in the prompt. We also experiment with \textbf{Zero-shot with RAG}, where we additionally include the top-$5$ relevant syllabi chunks retrieved in a process detailed below. As a sanity check, we include a \textbf{Search Baseline} where the answer to a question is simply the top-$1$ retrieved syllabus chunk, which reflects what a student would get using a simple keyword search (see Section~\ref{sec:appendix_search_baseline} in the Supplementary Material). 

In \textbf{SFT}, we fine-tune LLaMA-2-7B and LLaMA-2-13B on the \mbox{\textsc{SyllabusQA}} training set. In \textbf{SFT with RAG}, we additionally include the top-$5$ relevant syllabi chunks retrieved from the reference syllabus and construct the input prompt similar to zero-shot with RAG for a fair comparison. 

In \textbf{Retrieval-augmented Generation (RAG)}, we retrieve chunks in the course syllabi that are relevant to the question and include them in the input to LLMs. Although a shorter syllabus can be completely included in the prompt, we chose to retrieve only relevant chunks from the syllabus to allow for generalization to longer syllabi. Moreover, using long input contexts packed with irrelevant information can decrease model performance~\cite{liu2023lost}. In our RAG pipeline, we parse syllabi from raw PDFs to text files using Adobe Acrobat's PDF-to-text parser, which accurately parses varied text formats including tables, schedules, and lists. To accommodate the limited context length of LLMs, we chunk each syllabus to a maximum size of $1000$ characters per chunk and an overlap of $200$ characters between adjacent chunks to preserve context. We use BM25~\cite{robertson2009probabilistic}, a popular and effective ranking function, to retrieve the top-$5$ syllabus chunks to include in the input. 

We experiment with \textbf{Chain-of-Thought} prompting~\cite{wei2022chain}, leveraging the rich meta information of \mbox{\textsc{SyllabusQA}}. Specifically, we use an SFT with RAG approach on LLaMA-2-13B to first predict the question type, then generate reasoning steps written by the annotator for Single-hop and Multi-hop reasoning type questions only, before finally predicting the answer.

We also benchmark two additional approaches to provide more comparisons to the approaches above. The first is \textbf{GPT-4 with Retrieval Assistant} in a zero-shot prompting setting: we use the gpt-4-1106-preview model with the retrieval assistant from external files enabled, where we upload the raw PDF syllabus to GPT-4 and then prompt it to answer a question, along with system instructions explaining the task. Our prompt is shown in Section~\ref{sec:appendix_prompts} in the Supplementary Material. This approach provides us with an upper bound on performance of the zero-shot approach. The second is \textbf{Human Performance}, an estimated upper bound of QA performance on \mbox{\textsc{SyllabusQA}} for any approach. We simply calculate the agreement between answers written by the human expert and annotators on about $20\%$ of the test set. This measure underestimates human performance, though, since our annotators are not instructors who are familiar with the syllabi, making them susceptible to imperfect recall, especially on longer syllabi.


\subsection{Metrics}
\label{sec:metric}

We now detail the evaluation metrics we use for the QA task on \mbox{\textsc{SyllabusQA}}. We report surface textual similarity using traditional metrics like \textbf{ROUGE-L F1}~\cite{lin-2004-rouge} and recent ones like \textbf{BERTScore F1}~\cite{bert-score}. Since these metrics are more aligned with language style rather than the factuality of an answer, we also use several metrics to measure the \emph{factuality} of answers, which are more robust to surface textual features in answers than standard metrics.

\paragraph{Factuality Metrics}
We design a novel LLM-based (GPT-4) evaluation metric which we call \textbf{Fact-QA}, inspired by FActScore~\cite{factscore}, to evaluate the factuality of predicted answers. FActScore evaluates information precision by breaking the generated text into a series of atomic facts and then computing the percentage of atomic facts supported by a reference knowledge source, such as Wikipedia. The key modification we make in our adaptation is to use the annotator-written ground truth answer as the reference source. We then swap the predicted and ground truth answers to compute information recall.

To compute precision, we compute the proportion of facts in the predicted answer supported by the reference answer, denoted as Fact-QA Precision. To compute recall, we compute the proportion of facts in the reference answer supported by the predicted answer, denoted as Fact-QA Recall. Intuitively, if a predicted answer has high precision, it is less likely to contain incorrect or irrelevant facts, signifying limited LLM hallucination. If a predicted answer has high recall, it is more likely to have covered all relevant facts needed to answer the question. Both aspects are important since a desirable answer should cover all necessary facts but nothing irrelevant, which may distract students when they read the answer. We also aggregate these metrics and compute a Fact-QA F1 score. Please see Section~\ref{sec:appendix_metric} in the Supplementary Material for implementation details and our prompts.

We note that ideally, we should use the syllabus as the knowledge source to evaluate the factuality of an answer by providing it to GPT-4 for retrieval. However, due to the high cost associated with the OpenAI API, we use facts from the reference answer instead, which we found to be a high-quality summary of facts in the syllabus that are relevant to a question. Specifically, we examined $10\%$ of questions from the test set, where we compared the Fact-QA Precision and Recall scores obtained by providing the entire syllabus to GPT-4 vs.\ providing only the annotator-written answer. Results show a high correlation, with Pearson correlation coefficients of $0.8223$ and $0.7369$, respectively, which suggest that when computing the Fact-QA metric, using facts from the much shorter reference answer is a good proxy for using the full syllabus.

\paragraph{Qualitative Analysis of Fact-QA}
We (the authors) investigate the effectiveness of our Fact-QA metric by performing a qualitative analysis to judge whether GPT-4’s output is accurate on $5\%$ of the test set. We find that GPT-4's outputs are correct, partly flawed, and poor about $78\%$, $15\%$, and $7\%$ of the times, respectively. By examining GPT-4's errors in extracting facts and verifying claims, we find four major error types: 1) failures in logical reasoning, 2) overly granular atomic fact extraction, 3) arithmetic errors, and 4) incorrectly handling unanswerable questions (see Section~\ref{sec:appendix_metric_qualitative_analysis_factqa} in the Supplementary Material).

We (the authors) perform a human evaluation to examine the factual similarity between the expert annotator-written answers and the annotator-written ground-truth answers, on $5\%$ of the test set. We find moderate to high Pearson correlation coefficients of $0.7785$, $0.6602$, and $0.7735$ with Fact-QA Precision, Fact-QA Recall, and Fact-QA F1, respectively, indicating that Fact-QA is an effective metric (see Section~\ref{sec:appendix_metric_correlation_human} in the Supplementary Material).

\begin{table*}
\small
\centering
\begin{tabular}{p{0.35\linewidth}p{0.07\linewidth}p{0.09\linewidth}p{0.07\linewidth}p{0.08\linewidth}p{0.05\linewidth}}

\toprule

\multirow{3}{*}{Model} & \multicolumn{3}{c}{Factuality} &  \multicolumn{2}{c}{Surface Textual Similarity}\\
\cmidrule{2-6}
& Fact-QA F1 $\uparrow$ & Fact-QA Precision $\uparrow$ & Fact-QA Recall $\uparrow$ & ROUGE-L F1 $\uparrow$ & BERTScore F1 $\uparrow$\\

\midrule

\rowcolor{gray!21} \multicolumn{6}{c}{Zero-shot}\\
LLaMA-2-7B & $0.187$ & $0.131$ & $0.323$ & $0.086$ & $0.476$\\ 
LLaMA-2-13B & $0.192$ & $0.132$ & $0.353$ & $0.083$ & $0.475$\\
LLaMA-2-70B & $0.184$ & $0.129$ & $0.322$ & $0.090$ & $0.485$\\

\rowcolor{gray!21} \multicolumn{6}{c}{Zero-shot with Retrieval-augmented Generation}\\
Search Baseline & $0.267$ & $0.227$ & $0.324$ & $0.118$ & $0.477$\\
LLaMA-2-7B + RAG & $0.334$ & $0.228$ & $0.620$ & $0.111$ & $0.488$\\ 
LLaMA-2-13B + RAG & $0.332$ & $0.224$ & $0.635$ & $0.119$ & $0.497$\\
LLaMA-2-70B + RAG & $0.374$ & $0.261$ & $0.661$ & $0.146$ & $0.520$\\

\rowcolor{gray!21} \multicolumn{6}{c}{Supervised Fine-tuning}\\
LLaMA-2-7B + SFT & $0.280$ & $0.312$ & $0.253$ & $0.250$ & $0.585$\\ 
LLaMA-2-13B + SFT & $0.292$ & $0.312$ & $0.273$ & $0.269$ & $0.614$\\ 

\rowcolor{gray!21} \multicolumn{6}{c}{Supervised Finetuning with Retrieval-augmented Generation}\\
LLaMA-2-7B + SFT + RAG & $0.550$ & $\underline{0.579}$ & $0.524$ & $\underline{0.405}$ & $\underline{0.674}$\\ 
LLaMA-2-13B + SFT + RAG & $\underline{0.593}$ & $\mathbf{0.602}$ & $0.585$ & $\mathbf{0.429}$ & $\mathbf{0.702}$\\ 

\rowcolor{gray!21} \multicolumn{6}{c}{Supervised Finetuning with Retrieval-augmented Generation with Chain-of-Thought}\\
LLaMA-2-13B + SFT + RAG + CoT & $\mathbf{0.597}$ & $\mathbf{0.602}$ & $\underline{0.592}$ & $\mathbf{0.429}$ & $\mathbf{0.702}$\\ 

\rowcolor{gray!21} \multicolumn{6}{c}{GPT-4 with Retrieval Assistant}\\
GPT-4 + Retrieval Assistant & $0.561$ & $0.456$ & $\mathbf{0.728}$ & $0.260$ & $0.593$\\

\rowcolor{gray!21} \multicolumn{6}{c}{Human Performance}\\
Human & $0.685$ & $0.707$ & $0.664$ & $0.419$ & $0.684$\\ 

\bottomrule
\end{tabular}
\caption{Model performance on \mbox{\textsc{SyllabusQA}}. Search Baseline, GPT-4 with Retrieval Assistant and Human performance is reported on $20\%$ of the test set. Best (non-human) performance is in \textbf{bold} and the closest \underline{underlined}. 
}
\label{tab:results}
\end{table*}


\section{Results, Analysis, and Discussion}
\label{sec:results}

We report the performance on both surface textual similarity and factuality metrics for all approaches in Table~\ref{tab:results}, averaged over questions on the test set of \mbox{\textsc{SyllabusQA}}. We also provide stratified performance by question type in Table~\ref{tab:results_stratified}. We show qualitative example outputs across question types in Section~\ref{sec:appendix_qualitative_analysis} and an error analyses of failed cases in Section~\ref{sec:appendix_error_analysis} in the Supplementary Material.

\paragraph{Models are good at capturing surface textual similarity} 
The best-performing approach is LLaMA-2-13B with SFT combined with RAG and Chain-of-Thought (LLaMA-2-13B+SFT+RAG+CoT). This approach performs similarly to humans on surface-level textual similarity metrics. This observation suggests that training automated teaching assistant chatbots has some promise towards reducing human workload for course logistics-related QA. However, zero-shot approaches perform much worse than SFT on the same underlying LLM, which suggests that training on actual QA pairs is necessary to adapt to the diverse language styles among real instructors. We note that human performance is underestimated since we asked annotators to use diverse styles in their answers, which exaggerates disagreement~\cite{amidei-etal-2018-rethinking}.

\paragraph{Models fall behind humans on factuality}
We note that the performance of the best approach, LLaMA-2-13B+SFT+RAG+CoT, falls behind human performance on the factuality metrics and is especially lower on precision (by $10\%$), which indicates that LLMs are prone to hallucination. As one example, the best approach predicts ``Grading breakdown is exams (30\%), attendance (20\%), case analysis (25\%), and group project (25\%)'' while the reference answer is ``Grading breakdown is attendance (10\%), discussion (10\%), assignments (20\%), exams (40\%), and group project (20\%)''. The predicted answer scores highly on surface textual similarity but is inaccurate. See additional qualitative examples of LLM hallucination in Section~\ref{sec:appendix_error_analysis} in the Supplementary Material. We also found that LLMs perform worse on questions with implicit answers, especially reasoning-based ones, compared to questions with explicit answers (see Table~\ref{tab:results_stratified}).
These observations suggest that there is room for improvement before these approaches can be deployed in real-world educational settings, since course logistics questions can often be high-stakes especially when involving exams and grading, requiring answers to be factually accurate. Nonetheless, we believe that \mbox{\textsc{SyllabusQA}}, our FACT-QA metric, and these results serve as a useful benchmark for the research community moving forward.

\paragraph{Retrieval boosts performance}
We see that RAG provides a significant boost in LLM performance. In the zero-shot setting, LLaMA-2-70B combined with RAG shows an increase of $3.5\%$ on BERTScore F1 and $19\%$ on Fact-QA F1 than its non-RAG counterpart. Similarly, in the SFT setting, LLaMA-2-13B combined with RAG shows an increase of $8\%$ on BERTScore F1 and $30\%$ on Fact-QA F1 than its non-RAG counterpart. This improvement may even be more significant when using other retrieval methods such as dense passage retrieval~\cite{karpukhin-etal-2020-dense}, with the option of training the retriever using meta information of answer spans associated with explicit type questions, which we leave for future work.

\begin{table*}
\small
\centering
\begin{tabular}{p{0.2\linewidth} p{0.05\linewidth} p{0.14\linewidth} p{0.13\linewidth} p{0.17\linewidth} p{0.05\linewidth}}

\toprule
Category & Test Set $\%$ & GPT-4 + Retrieval Assistant & LLaMA-2-13B + SFT + RAG & LLaMA-2-13B \mbox{+ SFT} + RAG + CoT & Human \\
\midrule
\multicolumn{6}{c}{Stratified by Question Type}\\
\midrule
Yes/No & $14.32$ & $0.466$ & $0.811$ & $0.859$ & $0.802$\\
Single-factual & $14.32$ & $0.682$ & $0.600$ & $0.576$ & $0.833$\\
Multi-factual & $14.32$ & $0.551$ & $0.570$ & $0.552$ & $0.734$\\
Single-hop reasoning & $14.32$ & $0.496$ & $0.393$ & $0.365$ & $0.486$\\
Multi-hop reasoning & $14.32$ & $0.392$ & $0.261$ & $0.324$ & $0.374$\\
Summarization & $14.32$ & $0.387$ & $0.410$ & $0.407$ & $0.412$\\
Adversarial & $14.05$ & $0.416$ & $0.787$ & $0.774$ & $0.720$\\

\midrule
\multicolumn{6}{c}{Stratified by Source of Answer}\\
\midrule
Explicit & $42.97$ & $0.558$ & $0.660$ & $0.662$ & $0.786$\\
Implicit & $42.97$ & $0.423$ & $0.355$ & $0.365$ & $0.422$\\
Insufficient Info & $14.05$ & $0.416$ & $0.787$ & $0.774$ & $0.720$\\

\bottomrule
\end{tabular}
\caption{Stratified model performance by question type evaluated using the Fact-QA F1 metric on \mbox{\textsc{SyllabusQA}}. GPT-4 with Retrieval Assistant and Human performance is reported on $20\%$ of the test set.}
\label{tab:results_stratified}
\end{table*}

\paragraph{\mbox{\textsc{SyllabusQA}} is challenging even for GPT-4}
For the latest model from the GPT-4 family (gpt-4-1106-preview as of Dec 1, 2023) enabled with retrieval assistant where we upload the syllabus as a PDF file, we see that performance on textual similarity metrics is generally lower than that of SFT approaches which have been fine-tuned to adapt to the style of answers. GPT-4 performs similarly to humans on harder questions with implicit answers, especially reasoning-type ones (see Table~\ref{tab:results_stratified}). On Fact-QA Recall, GPT-4 surpasses human performance, which suggests that the retrieval assistant is highly effective. This result also indicates that human annotators may not retrieve relevant information perfectly, especially for long syllabi ($8$ pages on average). 

However, recall is only one side of the story. Compared to human performance on precision, GPT-4 with retrieval assistant underperforms, with a gap of $25\%$ in Fact-QA Precision and $12\%$ in Fact-QA F1 (possibly overestimated due to Yes/No type questions). This result indicates that while GPT-4 can retrieve relevant facts from the syllabus, its predicted answer is not always fully supported by the reference answer. Moreover, GPT-4 has a low Fact-QA F1 score on adversarial questions with insufficient supporting information in the syllabus (see Table~\ref{tab:results_stratified}). We show examples of GPT-4 hallucination in Section~\ref{sec:appendix_error_analysis} in the Supplementary Material. This factuality gap shows \mbox{\textsc{SyllabusQA}} is challenging even for state-of-the-art LLMs.

\paragraph{Chain-of-Thought improves answer factuality}
Chain-of-Thought with SFT and RAG leverages question type information, leading to improved performance on factuality metrics. Question type prediction (among 7 types) has an accuracy of $54\%$, much higher than random chance. If the question type is Single-hop or Multi-hop reasoning, the model then generates reasoning steps before predicting the answer. Investigating performance across question types (see Table~\ref{tab:results_stratified}), we see a significant increase in Fact-QA F1 performance on multi-hop reasoning questions. However, we see a small drop in performance on single-hop reasoning questions, possibly because these simpler questions do not benefit from intermediate reasoning steps.


\section{Future Work and Conclusions}

In this work, we introduced \mbox{\textsc{SyllabusQA}}, a diverse dataset consisting of real-world course syllabi and corresponding course logistics-related QA pairs. We also benchmarked the performance of several strong, LLM-based automated QA baselines on this dataset. Results show that fine-tuning and retrieval-augmented generation are helpful but there remains a significant gap between the performance of LLM-based approaches and that of humans on answer factuality. We make the \mbox{\textsc{SyllabusQA}} dataset, our code, and evaluation metrics publicly available to facilitate future work on automated teaching assistant development. There are plenty of avenues for future work. First, we can leverage question meta information to improve the performance of LLM-based QA approaches, and use overgenerate-and-rank to improve their robustness. Second, we can collect human labels on answer factuality and further improve QA approaches using methods such as direct policy optimization~\cite{tian2023fine}.


\section*{Limitations}

We run extensive automated and manual checks to control for artifacts in \mbox{\textsc{SyllabusQA}}. However, given the large-scale crowd-sourced annotation process, there could be artifacts present. Although we manually check a subset of \mbox{\textsc{SyllabusQA}} for acceptable data quality, we leave a full evaluation for future work. Further, due to the imperfect recall of human annotators on the reference syllabus, we found that the ground truth answers might not always be comprehensive or factually correct. \mbox{\textsc{SyllabusQA}} contains syllabi from courses taught in English. An important future direction is extending \mbox{\textsc{SyllabusQA}} to courses taught in other languages, making the benefits of automated teaching assistants accessible to students across different communities.


\section*{Ethical Considerations}
While we took several steps to ensure diversity in our dataset, we cannot guarantee diversity across student cultures, genders, or other demographics because we did not collect this information about annotators. As such, there is a risk of bias in our data, and we welcome future studies that investigate such risks in \mbox{\textsc{SyllabusQA}}. Our dataset is intended to be used for research purposes, and not to train systems deployed in classrooms without prior extensive risk mitigation. For real-world deployment of automated teaching assistants, the privacy of student data is critical, highlighting a drawback of using closed-source models like GPT-4. We hope that \mbox{\textsc{SyllabusQA}} serves as an important first step in introducing a dataset and evaluation benchmark for the development of open-source automated teaching assistant methods.


\section*{Acknowledgements}
The authors would like to thank all course instructors who agreed to share their syllabus and all \mbox{\textsc{SyllabusQA}} annotators. We thank the Learning Agency Lab for funding the data collection process, and Perpetual Baffour, Alex Franklin, Siyuan (Mei Mei) Li, Natalie Rambis, Jules King, Ulrich Boser, Hasnain Heickal, Jaewook Lee, Nischal Ashok Kumar, and Wanyong Feng for helpful discussions. We also thank the anonymous reviewers for their helpful comments. The authors are partially supported by the NSF under grant IIS-2202506.

\bibliography{custom}

\clearpage
\appendix


\section{Related Work}
\label{sec:appendix_rw}

In online courses, especially discussion forums, many prior works have attempted automated QA. A well-known example is Jill Watson \cite{goel2017using} and its later variants \cite{goel2021agent}. Jill Watson takes an expert systems approach using a knowledge base of human-authored QAs; when a question comes in, it searches through the knowledge base and determines whether an answer can be retrieved or generated. Other relevant works include the Curio SmartChat system \cite{ryan} and the Discussion-Bot system \cite{feng2006intelligent}. These works focus on retrieving relevant information and then use it in pre-defined interaction mechanisms with students. 
More recently, several works have approached this problem via automated, end-to-end QA systems. \citet{zylich2020exploring} used TF-IDF \cite{tfidf} to retrieve relevant documents and trained a recurrent neural network for end-to-end QA. While they also used a classifier to determine whether a question is answerable, they did not validate it on actual questions that are not answerable. In contrast, \mbox{\textsc{SyllabusQA}} contains such adversarial questions by design. The authors of a concurrent work \cite{hicke2023chata} employed LLMs with document retrieval to improve QA performance. While they also perform a human evaluation of the usefulness and accuracy of answers on a proprietary dataset, their focus is more on content-related QA rather than logistics-related QA.


\section{Dataset}
\label{sec:appendix_dataset_stats}

\paragraph{Dataset Collection}
We do not collect any personally identifiable information from crowd workers. We paid $18$-$20$ US dollars per HIT which on average took between $60$-$90$ minutes to complete, which is above the minimum wage in the United States. We collected explicit consent from each annotator before starting the task, and we informed them that their data would be used to train AI models for educational research. Our data collection process in this work is IRB-approved.

We take several steps to ensure that \textsc{SyllabusQA} is properly anonymized. We ensure that all names and identifying information, such as phone numbers, are removed from syllabus documents and file names. Additionally, we do not include the account IDs of crowd annotators in the dataset to ensure that answers are not linked to annotators. We additionally screen for offensive content in the data by manually checking a random subset of QA pairs written by each annotator and running automated heuristics like the profanity-check Python library.

\paragraph{Stratified Statistics by Question Type}
We show a breakdown of QA pairs in \mbox{\textsc{SyllabusQA}} by question type in Table~\ref{tab:syllabus_qa_stats_q_type}.

\begin{table}
\small
\centering
\begin{tabular}{p{0.4\linewidth} p{0.2\linewidth} p{0.2\linewidth}}

\toprule
Category & Count & Percentage $\%$ \\
\midrule
\multicolumn{3}{c}{Question Types}\\
\midrule
Yes/No & $725$ & $14.27$ \\
Single-factual & $725$ & $14.27$ \\
Multi-factual & $727$ & $14.31$ \\
Single-hop reasoning & $727$ & $14.31$ \\
Multi-hop reasoning & $727$ & $14.31$ \\
Summarization & $727$ & $14.31$ \\
Adversarial & $720$ & $14.17$ \\

\midrule
\multicolumn{3}{c}{Source of Answers}\\
\midrule
Explicit & $2177$ & $42.87$ \\
Implicit & $2181$ & $42.94$ \\
Insufficient Info & $720$ & $14.17$ \\

\bottomrule
\end{tabular}
\caption{Stratified statistics of the \mbox{\textsc{SyllabusQA}} dataset by question types defined in Section~\ref{subsection:q_type}.}
\label{tab:syllabus_qa_stats_q_type}
\end{table}

\paragraph{Comparison of Full Test Set vs $20\%$ Test Set}
As described in Sec~\ref{sec:annotationprocedure}, $20\%$ of questions were randomly sampled from the test set to compute agreement which was also used as an estimate of human performance. We also report GPT-4 with Retrieval Assistant performance on this same $20\%$ test set due to budgetary constraints. To investigate whether this $20\%$ test set is an accurate representation of the full test set, we first report statistics on the distribution of question types comparing the full test set to the $20\%$ test set in Table~\ref{tab:question_distribution_20_percent_test}. We observe that both sets have a sufficient number of questions from each question type. 

We additionally report model performance on the $20\%$ test set as well as the full test set in Table~\ref{tab:model_performance_20_percent_test}. We find similar model performance trends on both sets implying that the $20\%$ test set allows for a fair comparison.

\begin{table}
\small
\centering
\begin{tabular}{p{0.4\linewidth} p{0.2\linewidth} p{0.2\linewidth}}

\toprule
Category & Full Test Set & $20\%$ of Test Set \\
\midrule
\multicolumn{3}{c}{Question Types}\\
\midrule
Yes/No & $14.32\%$ & $16.50\%$ \\
Single-factual & $14.32\%$ & $22.50\%$ \\
Multi-factual & $14.32\%$ & $16\%$ \\
Single-hop reasoning & $14.32\%$ & $12.50\%$ \\
Multi-hop reasoning & $14.32\%$ & $7.50\%$ \\
Summarization & $14.32\%$ & $6.50\%$ \\
Adversarial & $14.05\%$ & $18.50\%$ \\

\midrule
\multicolumn{3}{c}{Source of Answers}\\
\midrule
Explicit & $42.97\%$ & $55\%$ \\
Implicit & $42.97\%$ & $26.50\%$ \\
Insufficient Info & $14.05\%$ & $18.50\%$ \\

\bottomrule
\end{tabular}
\caption{Distribution of question types in full test set vs $20\%$ test set.}
\label{tab:question_distribution_20_percent_test}
\end{table}

\paragraph{License}

We are releasing our dataset under the CC BY-NC-SA (Attribution-NonCommercial-ShareAlike) License, and clarify that the intended use is for research purposes.


\section{Analysis of Ground-truth Answers}
\label{sec:appendix_analysis_ground_truth_answers}

\subsection{Factuality Analysis of Ground-truth Answers}
We perform a factuality analysis of annotator-written ground-truth answers on two aspects: 1) precision and 2) recall. For precision, we test whether the information contained in the ground-truth answer is relevant to the question and supported by the syllabus. For recall, we test whether the major relevant information required to answer the question is contained in the ground-truth answer. An independent expert annotator manually verified a random sample of $10\%$ of the test set. For each QA pair, the expert annotator read the question, read the corresponding syllabus, and rated the ground-truth answer on precision and recall. Precision and recall were both individually rated using a three-point scale indicating poor, partial, and perfect scores.

Through this manual analysis performed by an independent expert annotator, we found that $82\%$ ($77\%$) of ground-truth answers have perfect precision (recall), $11\%$ ($16\%$) of ground-truth answers have partial precision (recall), and $7\%$ ($7\%$) of ground-truth answers have poor precision (recall). 

This error analysis highlights that human performance is underestimated. Annotators are susceptible to imperfect recall, especially on longer syllabi, reducing human performance. Ideally, human performance should be measured using the same professors who have written the course syllabi, thereby having close to perfect recall of its contents, leading to an improvement in human performance.

We performed a qualitative error analysis and found the following error types leading to less-than-perfect precision/recall in ground-truth answers: 1) imperfect annotator 
recall/human error, 2) ambiguous syllabus, and 3) ambiguous answers. We provide examples for each of the error types in Table~\ref{tab:error_analysis_ground_truth_answers}.

\begin{table*}
\small
\centering
\scalebox{1}{
\begin{tabular}{p{.3\linewidth}p{.6\linewidth}}

\toprule

\multicolumn{2}{c}{Error Type: Imperfect Annotator Recall/Human Error}\\
\midrule
Question & What was the punishment for academic dishonesty?\\
Ground-truth Answer & No/insufficient information\\
Remark & The syllabus states ``Appropriate sanctions may be imposed on any student''. This information is contained within a large paragraph at the end.\\

\midrule
\multicolumn{2}{c}{Error Type: Ambiguous Syllabus}\\
\midrule
Question & When are the TA Help Sessions?\\
Ground-truth Answer & Thursdays over Zoom. Additionally Fridays when homework is due. Time TBD.\\
Remark & The syllabus states the time as ``Thursday 5:00-6:00 pm and Friday 4:00-5:00 pm'' in one place while also stating the time as ``TBD'' in another place.\\

\midrule
\multicolumn{2}{c}{Error Type: Ambiguous Answer}\\
\midrule
Question & Will the two mid-term exams in this course cover cumulative material or only the material covered since the previous exam?\\
Ground-truth Answer & No\\
Remark & The QA pair is incorrectly formulated as a yes/no question type instead of a single factual leading to ambiguity in the answer.\\

\bottomrule
\end{tabular}
}
\caption{Examples of ground-truth answers in \mbox{\textsc{SyllabusQA}} that exhibit partial factuality covering major error types found in a manual qualitative analysis of ground-truth answers by an independent expert annotator.}
\label{tab:error_analysis_ground_truth_answers}
\end{table*}


\subsection{Correlation of Ground-truth Answers with Syllabus Text}

For an illustration of how ground-truth answers to implicit vs explicit question types differ, we provide an example QA pair for multi-hop reasoning (implicit) and multi-factual (explicit) type questions in Table~\ref{tab:correlation_ground_truth_answers_syllabus}. We observe that implicit question types have answers that can be very different from the syllabus text spans often requiring reasoning and summarization, while for explicit question types, the answers are more correlated to the text spans of the syllabus although written in an open-ended natural style.

\begin{table*}
\small
\centering
\scalebox{1}{
\begin{tabular}{p{.3\linewidth}p{.6\linewidth}}

\toprule

\multicolumn{2}{c}{Question Type: Implicit (Multi-hop Reasoning)}\\
\midrule
Question & Will I get a zero if I miss a quiz because I am sick?\\
Relevant Syllabus Text Span 1 & There are no extensions for quizzes, but instead the lowest (1) quiz score will be dropped at the end of the semester, to accommodate for any reason a quiz might be missed.\\
Relevant Syllabus Text Span 2 & If you become ill it is important that you test and ensure that you have a negative test before coming to class. In such a case you should contact your instructor (for homework and exams) and your discussion TA (for quizzes) to make alternative arrangements.\\
Ground-truth Answer & You should contact your TA for individual accommodations, class policy states that quizzes cannot be rescheduled but your lowest score will be dropped so missing a quiz should have no impact on your course grade.\\

\midrule
\multicolumn{2}{c}{Question Type: Explicit (Multi-factual)}\\
\midrule
Question & When is my paper proposal due and how much is it worth?\\
Relevant Syllabus Text Span 1 & Reform Paper Proposal Due 9:00PM March 24\\
Relevant Syllabus Text Span 2 & Reform Paper (5\% Proposal,\\
Ground-truth Answer & The paper proposal is worth 5\% of your grade and is due by 9pm on March 24th\\

\bottomrule
\end{tabular}
}
\caption{Correlation of ground-truth answers with relevant text spans from the source syllabus for implicit vs explicit question types.}
\label{tab:correlation_ground_truth_answers_syllabus}
\end{table*}


\section{FACT-QA Metric}
\label{sec:appendix_metric}

\subsection{Implementation Details and Prompts}
For the Fact-QA metric, we use GPT-4, specifically gpt-4-1106-preview, as the underlying LLM for Fact-QA. In our prompts, we do not reveal to GPT-4 the roles of the answers and instead refer to them as ``answer 1'' and ``answer 2''. Given a question and two answers, we have GPT-4 extract all atomic facts from answer 1 and compute how many are supported by answer 2. We prompt GPT-4 in this way twice for each predicted answer, setting answer 1 to the predicted answer and answer 2 to the reference answer for Fact-QA Precision and vice versa for Fact-QA Recall. We provide the prompt and example output for Fact-QA in Table~\ref{tab:metric-prompt}.

\subsection{Qualitative Analysis of Fact-QA Metric}
\label{sec:appendix_metric_qualitative_analysis_factqa}

We investigate the effectiveness of our Fact-QA metric by performing a qualitative analysis of GPT-4’s outputs when evaluating the expert annotator-written answers compared against the ground-truth answers on $5\%$ of the test set. We find that for around $78\%$ of evaluations, GPT-4 executes the task correctly and assigns scores that are a good representation of the factual similarity between the answers. For around $15\%$ of evaluations, GPT-4 had minor issues with Fact-QA being a satisfactory representation of the factual quality of the answer. For the remaining $7\%$ of evaluations, either GPT-4 had major issues or the metric is not a good reflection of answer quality.

Among the strengths of using Fact-QA, GPT-4 is effective at handling answers with varied styles, is capable of extracting atomic facts from the first answer which can be suitably compared to the second answer, and exhibits excellent (implicit) reasoning to determine which facts are supported by the second answer. Additionally, we find that hallucinations are surprisingly rare, possibly due to the straightforward nature of the Fact-QA task of comparing two answers.

For a qualitative error analysis, we additionally examine the subset of partly flawed ($15\%$) and poor ($7\%$) cases where GPT-4 makes errors in either extracting atomic facts from the first answer or verifying claims stated in these facts by the second answer. First, we observe that GPT-4 occasionally fails with logical reasoning. For example, GPT-4 determined that a fact claiming a grade would be a zero was supported by an answer claiming a grade would be hurt, when only the converse is true. Second, we observe that GPT-4 at times tends to extract multiple facts by breaking a single sentence into $3$ or more facts. This granular extraction can cause scores to be overly low when some information is not supported by the second answer. Third, we observe that GPT-4 often fails to understand the meaning of ``No/Insufficient Information'', inferring this to be the same as a direct ``No'' answer to a question. Fourth, we observe that GPT-4 at times struggles with arithmetic to determine if facts are supported. For example, GPT-4 fails to identify that $87\%$ and $26/30$ are roughly equal. We note that many of these issues could be fixed with further prompt engineering, which we leave for future work.

To conclude our analysis of the Fact-QA Metric, we present several observations that could lead to improvements in the metric for future work. First, we note that a fact unsupported by the second answer is treated equally to a fact directly contradicted by the second answer. An additional penalty for contradicting information would be helpful. 
Second, we note that for some answers not relevant to the question, GPT-4 still attempts to interpret them. Automatically assigning a score of $0$ to irrelevant answers as an initial screening might help.

\subsection{Correlation of Fact-QA Metric with Human Evaluation}
\label{sec:appendix_metric_correlation_human}

We (the authors) additionally perform a human evaluation to examine the factual similarity between the expert annotator-written answers and the annotator-written ground-truth answers, on $5\%$ of the test set. We set up our human evaluation imitating the Fact-QA evaluation process. For the manual analysis, we use only the question, the predicted answer, and the ground-truth answer, matching Fact-QA. For precision, we assess if the factual information contained in the predicted answer is supported by the ground-truth answer. For recall, we assess if the factual information contained in the ground-truth answer is supported by the predicted answer. We assign a human evaluation precision/recall score of $0$, $0.5$, and $1$ for poor, partial, and complete precision/recall, respectively. We find moderate to high Pearson correlation coefficients of $0.7785$, $0.6602$, and $0.7735$ with Fact-QA Precision, Fact-QA Recall, and Fact-QA F1, respectively, indicating that Fact-QA is an effective metric.

\subsection{Average Number of Atomic Facts in Ground-truth Answers}
To illustrate the role of atomic facts in the Fact-QA metric, we report statistics on the number of atomic facts per ground-truth answer in the test set, stratified by question type in Table~\ref{tab:num_facts_answer}.

\begin{table}
\small
\centering
\begin{tabular}{p{0.35\linewidth} p{0.12\linewidth} p{0.08\linewidth} p{0.08\linewidth} p{0.08\linewidth}}

\toprule
Question Type & Average & Std Dev & Min & Max \\
\midrule
Yes/No & $1.03$ & $0.18$ & $1$ & $2$\\
Single-factual & $1.59$ & $0.95$ & $1$ & $6$\\
Multi-factual & $4.36$ & $2.21$ & $1$ & $12$\\
Single-hop reasoning & $2.2$ & $1.05$ & $1$ & $6$\\
Multi-hop reasoning & $3.34$ & $1.74$ & $0$ & $9$\\
Summarization & $4.8$ & $2.19$ & $0$ & $11$\\
Adversarial & $1.06$ & $0.25$ & $1$ & $2$\\

\midrule
Overall & $2.63$ & $2.05$ & $0$ & $12$\\

\bottomrule
\end{tabular}
\caption{Stratified statistics on the number of atomic facts automatically extracted per ground-truth answer by our Fact-QA metric.}
\label{tab:num_facts_answer}
\end{table}

\subsection{Simple Example Illustrating the Fact-QA Metric}
We provide a simple example to illustrate our Fact-QA metric. In the example shown in Table~\ref{tab:simple_example_factqa}, Answer 1 is the ground-truth answer, and Answer 2 is the predicted answer. We get a Fact-QA Recall score of 1/2. To find Fact-QA Precision, we swap Answer 1 and Answer 2 to get a Fact-QA Precision score of 1.

\begin{table*}
\small
\centering
\begin{tabular}{p{0.2\linewidth} p{0.7\linewidth}}
\toprule
Prompt & <instructions>\\
&Question: What date do the seminar meetings begin?\\
&Answer 1: The seminar meetings will begin on February 10 from 4:30-6:00pm.\\
&Answer 2: Friday, February 10 is the first day of the seminar.\\

\midrule
GPT4-Response & 1. Atomic claims made by Answer 1:\\
&\quad The seminar meetings will begin on February 10.\\
&\quad The seminar meetings will be from 4:30-6:00pm.\\\\
&2. Claims supported by Answer 2:\\
&\quad The seminar meetings will begin on February 10.\\\\
&3. Score: 1/2\\

\bottomrule
\end{tabular}
\caption{Simple example illustrating our Fact-QA metric.}
\label{tab:simple_example_factqa}
\end{table*}


\section{Search Baseline as a Sanity Check}
\label{sec:appendix_search_baseline}

We implement a simple rule-based keyword search baseline as a useful sanity check for model performance on SyllabusQA. The search baseline imitates a student searching for an answer within the syllabus using keywords from the question. We chunk syllabus text documents into chunks of $200$ characters with an overlap of 50 characters between adjacent chunks to preserve context. We chose a chunk size of 200 characters since this roughly corresponds to the mean number of tokens in an answer in the train set of SyllabusQA ($30.3$ tokens/answer), and the answer length provides a balance between recall and precision. We retrieve the top-1 chunk using BM-25, a strong and effective retrieval function simulating a manual keyword search performed by a student. We simply output this retrieved top-1 chunk without any modification as the predicted answer.

The search baseline is competitive with Zero-shot LLaMA with RAG on textual similarity and Fact-QA Precision as seen in Table~\ref{tab:results}. Since Zero-shot LLaMA is not fine-tuned on SyllabusQA, the answers generated are sometimes in a different chatbot style, which reduces textual similarity scores. In contrast, the search baseline outputs an unfiltered text snippet from the syllabus. On Fact-QA precision, the choice of using only the top-1 retrieved snippet ensures higher precision but reduces recall since Zero-shot LLaMA with RAG synthesizes information from top-5 retrieved snippets.

While serving as a useful sanity check, the search baseline has major drawbacks compared to our models including 1) not being able to answer questions in a conversational style, 2) not being able to reason about facts spread across retrieved snippets to answer complex questions, 3) not being able to summarize multiple facts spread across the syllabus in a concise answer, and 4) not being able to identify when an answer is unanswerable from the retrieved snippets.

We illustrate through qualitative examples in Table~\ref{tab:error_analysis_search_baseline} questions, especially of type implicit, on which the Search Baseline performs poorly including multi-reasoning and summarization-type questions. We show the top-1 syllabus snippet searched/retrieved by BM-25 as the predicted answer.

\begin{table*}
\small
\centering
\scalebox{1}{
\begin{tabular}{p{.3\linewidth}p{.6\linewidth}}

\toprule

\multicolumn{2}{c}{Implicit (multi-reasoning) Question Type}\\
\midrule
Question & What's the maximum grade I can get on the forum posts if I don't work on them during exam weeks?\\
Search Baseline Output & reminders during class. You can email me with your questions (I aim to reply within 24 hours on weekdays, and 48 hours on weekends), or post your questions to the Question and Answer (Q\&A) forum on\\
Ground-truth Reasoning Step 1 & Weeks 4, 7, 11, and 14 are exam weeks.\\
Ground-truth Reasoning Step 2 & Weeks 4 and 7 allow a possible 3 points each for the forums, and weeks 11 and 14 allow a possible 4 points each for the forums, totaling 14 points.\\
Ground-truth Reasoning step 3 & The maximum allowed points from forum points and responses is 45, so you could get a max of 31 points.\\
Ground-truth Reasoning step 4 & The forum is graded with a maximum of 30 points, so you could get 100\%.\\
Ground-truth Answer: & You can still get $100\%$ credit for the forums if you didn't interact with them on exam weeks.\\

\midrule
\multicolumn{2}{c}{Implicit (summarization) Question Type}\\
\midrule
Question & What is the grading breakdown for this course?\\
Search Baseline Output & voluntary. Choosing not to participate in extra credit research will NOT have any negative impact on your course grade.\\
Ground-truth Answer & The grading breakdown is as follows: Class participation is worth $10\%$, weekly discussion on Blackboard is worth $10\%$, Individual assignments are worth $20\%$ (two - $10\%$ each), Midterm exams are worth $40\%$ (two $20\%$ each), and Group projects are worth $20\%$ (two - the assessment report is worth $15\%$ and the social venture pitch is wort $5\%$).\\

\bottomrule
\end{tabular}
}
\caption{Qualitative examples of questions, especially of type implicit, on which the Search Baseline performs poorly including multi-reasoning and summarization-type questions.}
\label{tab:error_analysis_search_baseline}
\end{table*}


\section{Experimental Setup}
\label{sec:appendix_params}

For SFT with LLaMA models, we use the AdamW optimizer with a batch size of 8, a learning rate of $1e$-$4$, and a cosine learning rate scheduler. We warm up for $3\%$ of training steps and use gradient clipping for training stability. 
We use the Parameter Efficient Fine-Tuning (PEFT) library from HuggingFace~\cite{wolf-etal-2020-transformers} to load LLaMA models and train via low-rank adaptation (LoRA)~\cite{hu2022lora} ($\text{LoRA }\alpha=32, \text{LoRA }r=16, \text{LoRA dropout}=0.05$). We use full precision for LLaMA-7B and LLaMA-13B models and 8-bit precision for inference on LLaMA-70B. We fine-tune for $3$ epochs with early stopping on the validation set on a single A$100$ $80$GB GPU, with each epoch taking up to 4 hours. 
For RAG, we use the Okapi BM25~\cite{robertson2009probabilistic} ranking function to retrieve the top-$5$ syllabus chunks with default parameters ($k_1=1.5, b=0.75$). For generation, we use nucleus sampling with top-k ($p=0.95, k=50$). Wherever possible, we use standard hyperparameters and do not do extensive parameter tuning since we aim to benchmark the performance of various baselines on \mbox{\textsc{SyllabusQA}}.
Our goal is not to develop the best QA model in this work. Due to high computational and Open AI API cost, we report performance on one run of model baselines. For ROUGE we use the rouge-score library with Porter stemmer enabled, and for BERTScore we use the bert-score library with microsoft/deberta-xlarge-mnli as the underlying model.
We additionally note that we used GitHub Copilot minimally in the writing of our code. All software we use in the development of this work is open source. We are consistent with the terms and intended use of all software and with the OpenAI API.


\section{Additional Results}
\label{sec:appendix_results}

\subsection{Model Performance on Full Test Set vs \texorpdfstring{$20\%$}{20\%} Test Set}

We reported GPT-4 with Retrieval Assistant and human performance on the $20\%$ test set in our main results in Table~\ref{tab:results}. To investigate whether this is fair, we compare the performance of all models on the full test vs $20\%$ test set in Table~\ref{tab:model_performance_20_percent_test} using BERTScore F1 as the evaluation metric. We find similar model performance trends on both sets implying that the $20\%$ test set allows for a fair comparison.

\begin{table*}
\small
\centering
\begin{tabular}{p{0.45\linewidth}p{0.2\linewidth}p{0.2\linewidth}}

\toprule
Model & Full Test Set & $20\%$ of Test Set\\

\midrule

\rowcolor{gray!21} \multicolumn{3}{c}{Zero-shot}\\
LLaMA-2-7B & $0.476$ & $0.472$\\ 
LLaMA-2-13B & $0.475$ & $0.468$\\ 
LLaMA-2-70B & $0.485$ & $0.476$\\ 

\rowcolor{gray!21} \multicolumn{3}{c}{Zero-shot with Retrieval-augmented Generation}\\
LLaMA-2-7B + RAG & $0.488$ & $0.481$\\  
LLaMA-2-13B + RAG & $0.497$ & $0.492$\\ 
LLaMA-2-70B + RAG & $0.52$ & $0.511$\\ 

\rowcolor{gray!21} \multicolumn{3}{c}{Supervised Fine-tuning}\\
LLaMA-2-7B + SFT & $0.585$ & $0.578$\\ 
LLaMA-2-13B + SFT & $0.614$ & $0.588$\\ 

\rowcolor{gray!21} \multicolumn{3}{c}{Supervised Finetuning with Retrieval-augmented Generation}\\
LLaMA-2-7B + SFT + RAG & $0.674$ & $0.641$\\ 
LLaMA-2-13B + SFT + RAG & $0.702$ & $0.689$\\ 

\rowcolor{gray!21} \multicolumn{3}{c}{Supervised Finetuning with Retrieval-augmented Generation with Chain-of-Thought}\\
LLaMA-2-13B + SFT + RAG + CoT & $0.702$ & $0.697$\\ 

\rowcolor{gray!21} \multicolumn{3}{c}{GPT-4 with Retrieval Assistant}\\
GPT-4 + Retrieval Assistant & NA & $0.593$\\ 

\rowcolor{gray!21} \multicolumn{3}{c}{Human Performance}\\
Human & NA & $0.684$\\ 

\bottomrule
\end{tabular}
\caption{Model performance on full test set vs $20\%$ test set using BERTScore F1. We find similar performance trends on both sets implying that the $20\%$ test set allows for a fair comparison.}
\label{tab:model_performance_20_percent_test}
\end{table*}


\section{Qualitative Analysis}
\label{sec:appendix_qualitative_analysis}
We show example qualitative outputs from the best-performing models for questions from each type in Table~\ref{tab:qualitative_analysis}.

\begin{table*}
\small
\centering
\scalebox{.8}{
\begin{tabular}{p{.3\linewidth}p{.7\linewidth}}

\toprule

\multicolumn{2}{c}{Yes/No}\\
\midrule
Question & Do I need to come to class in person?\\
Ground-truth answer & No\\
LLaMA-2-13B + SFT + RAG + CoT & No, you do not need to come to class in person since classes are held online.\\
GPT-4 + Retrieval Assistant & Though attendance at the lectures is not required, it is strongly encouraged to attend each lecture as material that is not discussed in the course materials (and will appear on the exams) will be covered. However, if a student misses a lecture session, they can view the lecture on the course Moodle page through Echo360.\\

\midrule
\multicolumn{2}{c}{Single-factual}\\
\midrule
Question & How many credits does this course provide?\\
Ground-truth answer & This course will provide 3 credits.\\
LLaMA-2-13B + SFT + RAG + CoT & This course provides 3 credits.\\
GPT-4 + Retrieval Assistant & The course provides 3 credit hours.\\

\midrule
\multicolumn{2}{c}{Multi-factual}\\
\midrule
Question & How will I know whether my prerequisite course performance is sufficient enough to enroll in this course?\\
Ground-truth answer & If you received a B or higher in COMPSCI 311, you should be able to succeed in the course. Additionally, if you demonstrated sufficient ability in complexity theory, number theory, abstract algebra, and combinatorics--even if you did not take COMPSCI 311--the instructor may waive the requirement.\\
LLaMA-2-13B + SFT + RAG + CoT & You should have earned a B or higher in the prerequisite course COMPSCI 311. If you earned less than a B, it is recommended that you reconsider taking the class. Additionally, the instructor may waive the formal prerequisite if you have taken courses other than COMPSCI 311 that demonstrate mathematical maturity. Please consult the instructor if you are not sure you have the right background.\\
GPT-4 + Retrieval Assistant & To determine if your prerequisite course performance is sufficient to enroll in the course, you should reflect on your performance in COMPSCI 311. If you earned less than a B in COMPSCI 311, it is recommended that you reconsider taking the class. However, mathematical maturity is a more fundamental prerequisite, and if you have taken other courses that demonstrate this maturity (such as complexity theory, number theory, abstract algebra, or combinatorics), the instructor may waive the formal prerequisite. If you are unsure whether you have the right background for the course, you should consult the instructor.\\

\midrule
\multicolumn{2}{c}{Single-hop reasoning}\\
\midrule
Question & I haven't taken Physics 605. Do I satisfy this course's prerequisites?\\
Ground-truth answer & No, you do not satisfy the prerequisites since completion of both Physics 601 and Physics 605 are required in order to take this class.\\
LLaMA-2-13B + SFT + RAG + CoT & No, PHYSICS 605 is a prerequisite for this course.\\
GPT-4 + Retrieval Assistant & No, you do not satisfy the course prerequisites as Physics 605 is a required prerequisite.\\

\midrule
\multicolumn{2}{c}{Multi-hop reasoning}\\
\midrule
Question & Will I get a zero if I miss a quiz because I am sick?\\
Ground-truth answer & You should contact your TA for individual accommodations, class policy states that quizzes cannot be rescheduled but your lowest score will be dropped so missing a quiz should have no impact on your course grade. \\
LLaMA-2-13B + SFT + RAG + CoT & Quizzes are worth $5\%$ of the final grade. If a student misses 1 quiz, the lowest quiz score will be dropped at the end of the semester.\\
GPT-4 + Retrieval Assistant & No, you will not automatically get a zero if you miss a quiz because you are sick. The course policy states that there are no extensions for quizzes; however, the lowest quiz score will be dropped at the end of the semester to accommodate for any reason a quiz might be missed.\\

\midrule
\multicolumn{2}{c}{Summarization}\\
\midrule
Question & How does the grade for discussion group does work?\\
Ground-truth answer & The discussion is done on regular basis and assignment will be given which they need to work alone and based on that grades will be given.\\
LLaMA-2-13B + SFT + RAG + CoT & Discussion group work is graded. At the end of every discussion, there will be a graded assignment that you will have to work on alone. It will be either the same as or similar to the work you do with a group at the beginning of discussion and will be worth half of the discussion grade. If you must miss your discussion, make it up during one of the other sections (they are listed on moodle). However, you MUST make sure both your TA and the TA of the section you do the make up in, know that you are doing this.\\
GPT-4 + Retrieval Assistant & The grade for the discussion group works as follows: Your group work in the discussion is graded. At the end of every discussion, there will be a graded assignment that you must complete individually. This individual assignment will be either the same as or similar to the work you initially did with your group. It is worth half of the discussion grade.\\

\midrule
\multicolumn{2}{c}{Adversarial}\\
\midrule
Question & What are the exact dates of all exams and assignments?\\
Ground-truth answer & No/insufficient information\\
LLaMA-2-13B + SFT + RAG + CoT & No/insufficient information\\
GPT-4 + Retrieval Assistant & No/insufficient information\\

\bottomrule
\end{tabular}
}
\caption{Example qualitative predicted answers from the best-performing models for questions from each type in \mbox{\textsc{SyllabusQA}}.}
\label{tab:qualitative_analysis}
\end{table*}


\section{Qualitative Error Analysis}
\label{sec:appendix_error_analysis}
We show examples of factually incorrect answers indicating model hallucination, or answers with additional irrelevant information, from the best-performing model, LLaMA-2-13B + SFT + RAG + CoT in Table~\ref{tab:error_analysis_llama}, and from GPT-4 with Retrieval Assistant in Table~\ref{tab:error_analysis_gpt4}.

\begin{table*}
\small
\centering
\scalebox{1}{
\begin{tabular}{p{.3\linewidth}p{.6\linewidth}}

\toprule

\multicolumn{2}{c}{Yes/No}\\
\midrule
Question & Are the lectures for this course in person?\\
Ground-truth answer & Yes\\
LLaMA-2-13B + SFT + RAG + CoT & No\\

\midrule
\multicolumn{2}{c}{Single-factual}\\
\midrule
Question & What are the points for the final? \\
Ground-truth answer & 200 points\\
LLaMA-2-13B + SFT + RAG + CoT & No/insufficient information\\

\midrule
\multicolumn{2}{c}{Multi-factual}\\
\midrule
Question & How much of the final grade is going to be comprised of exams?\\
Ground-truth answer & Exams are going to account for $60\%$ of the final grade. $30\%$ will be the in class midterm while the other $30\%$ is going to be the final exam.\\
LLaMA-2-13B + SFT + RAG + CoT & The final grade is going to be comprised of $30\%$ of exams.\\

\midrule
\multicolumn{2}{c}{Single-hop reasoning}\\
\midrule
Question & This is my first semester in this career path, should I take other courses before this one?\\
Ground-truth answer & Since this course has not pre-requisites, you will be fine taking it at anytime before any course that requires this one as a pre-requisite.\\
LLaMA-2-13B + SFT + RAG + CoT & Yes, it's a good idea to take other courses before this one.\\

\midrule
\multicolumn{2}{c}{Multi-hop reasoning}\\
\midrule
Question & If I forget to submit a quiz, can I still get a perfect score on my quiz grade?\\
Ground-truth answer & Yes, the two lowest quiz scores are dropped, so it is possible to miss a reading quiz and still maintain a perfect score.\\
LLaMA-2-13B + SFT + RAG + CoT & No, if you forget to submit a quiz, you will not be able to get a perfect score on your quiz grade because the lowest possible score is $0\%$.\\

\midrule
\multicolumn{2}{c}{Summarization}\\
\midrule
Question & What is the grading breakdown for this course?\\
Ground-truth answer & The grading breakdown is as follows: Class participation is worth $10\%$, weekly discussion on Blackboard is worth $10\%$, Individual assignments are worth $20\%$ (two - $10\%$ each), Midterm exams are worth $40\%$ (two $20\%$ each), and Group projects are worth $20\%$ (two - the assessment report is worth $15\%$ and the social venture pitch is wort $5\%$). \\
LLaMA-2-13B + SFT + RAG + CoT & The grading breakdown for this course is as follows: 1) Exams (2) $30\%$ 2) Class Participation/Attendance $20\%$ 3) Case Analysis $25\%$ 4) Social Venture Profile $25\%$\\

\midrule
\multicolumn{2}{c}{Adversarial}\\
\midrule
Question & What technology is present in the classroom during lectures?\\
Ground-truth answer & No/insufficient information\\
LLaMA-2-13B + SFT + RAG + CoT & The classroom is equipped with Zoom, which is used for lectures, as well as a projector and screen for presentations.\\

\bottomrule
\end{tabular}
}
\caption{Examples of factually incorrect predicted answers by the best-performing model, LLaMA-2-13B + SFT + RAG + CoT, on questions from each type in \mbox{\textsc{SyllabusQA}}.}
\label{tab:error_analysis_llama}
\end{table*}

\begin{table*}
\small
\centering
\scalebox{1}{
\begin{tabular}{p{.3\linewidth}p{.6\linewidth}}

\toprule

\multicolumn{2}{c}{Yes/No}\\
\midrule
Question & Do we need to be familiar with programming before taking this course?\\
Ground-truth answer & Yes\\
GPT-4 + Retrieval Assistant & No/insufficient information.\\

\midrule
\multicolumn{2}{c}{Single-factual}\\
\midrule
Question & Where can I find the required reading?\\
Ground-truth answer & You can find the required reading on Moodle. \\
GPT-4 + Retrieval Assistant & There is no required textbook for this class. Required readings or viewings and an ungraded reading guide will be available each week on Moodle, typically posted on Wednesday, and you will have until Friday at midnight to complete them.\\

\midrule
\multicolumn{2}{c}{Multi-factual}\\
\midrule
Question & When will we learn about enzymes?\\
Ground-truth answer & Week 4 (Feb. 27-Mar. 5), Week 5 \& Spring Break (Mar. 6-19),  and Week 6 (Mar. 20-26)\\
GPT-4 + Retrieval Assistant & You will learn about enzymes during Week 5, which includes topics on how enzymes work and the dependence of enzyme function on amino acid properties.\\

\midrule
\multicolumn{2}{c}{Single-hop reasoning}\\
\midrule
Question & How many classes will focus on plotting?\\
Ground-truth answer & There are three sessions devoted to learning plotting.\\
GPT-4 + Retrieval Assistant & The course will focus on plotting during two classes: one on "More plotting" and another on "Interactive plotting".\\

\midrule
\multicolumn{2}{c}{Multi-hop reasoning}\\
\midrule
Question & If I skip the Team Social Venture pitch project, can I still get an A?\\
Ground-truth answer & Yes, it is still possible to to get an A even if the Team Social Venture project is not completed.\\
GPT-4 + Retrieval Assistant & No, the Team Social Venture Pitch project is worth $5\%$ of your total grade. Since an `A' grade requires a $93\%$ or higher, missing the project would not allow you to achieve an `A'.\\

\midrule
\multicolumn{2}{c}{Summarization}\\
\midrule
Question & What is the percentage of the amount of your final grade that comes from student teacher seminar?\\
Ground-truth answer & The portion of your final grade that comes from student teacher seminar will be weighted as follows: Midpoints reports (2)- $15\%$, Final reports (2) -$15\%$, Weekly assignments - $30\%$, resume, Resume - $20\%$, Sample cover letter -$20\%$.\\
GPT-4 + Retrieval Assistant & The percentage of the final grade that comes from the student teacher seminar is 25\%.\\

\midrule
\multicolumn{2}{c}{Adversarial}\\
\midrule
Question & What is the required reading for this course?\\
Ground-truth answer & No/insufficient information\\
GPT-4 + Retrieval Assistant & The required reading for this course includes the following additional references for students to expand their knowledge about the subject: 1. L. Reimer ‚ Scanning Electron Microscope, 2nd Ed., Springer-Verlag 1998 2. D. Koningsberger \& R. Prins‚ X-ray Absorption Spectroscopy: Principles, Applications, and Techniques of EXAFS, SEXAFS, and XANES,‚Wiley 1988 3. John F. Moulder‚ Handbook of X-ray Photoelectron Spectroscopy, Perkin-Elmer Corp. 1992 4. Frans D. Tichelaar‚ Transmission Electron Microscopy as Nanolab, Wiley 2012.\\

\bottomrule
\end{tabular}
}
\caption{Examples of predicted answers that are either fatually incorrect or contain additional irrelevant information, by GPT-4 with Retrieval Assistant, on questions from each type in \mbox{\textsc{SyllabusQA}}.}
\label{tab:error_analysis_gpt4}
\end{table*}


\section{Prompts}
\label{sec:appendix_prompts}

We list all prompts for reproducibility.

\subsection{Prompts for LLaMA Models}

For models from the LLaMA-2 family, we match their pre-training prompt style and use the following format:

\noindent\fbox{%
    \parbox{\linewidth}{%
        $<$s$>$[INST] $<<$SYS$>>$\\
        You are a teaching assistant. Answer questions from students on course logistics. $<<$/SYS$>>$\\
        \{RAG context text for RAG baselines\}\\
        \#\#\# The question is: \{question text\}\\
        \#\#\# The answer is: [/INST]\\
    }%
}

The RAG context text for RAG baselines is constructed as follows:

\noindent\fbox{%
    \parbox{\linewidth}{%
        Here are snippets from the course syllabus which could be helpful in answering the question:\\
        \#\#\# Snippet 1: \{snippet 1 text\}\\
        \#\#\# Snippet 2: \{snippet 2 text\}\\
        \#\#\# Snippet 3: \{snippet 3 text\}\\
        \#\#\# Snippet 4: \{snippet 4 text\}\\
        \#\#\# Snippet 5: \{snippet 5 text\}\\
        The relevant snippets from the course syllabus were added above.
    }%
}

\subsection{Prompts for LLaMA Models with Chain-of-Thought Prompting}
We combine Chain-of-Thought prompting using the following prompt style:

\noindent\fbox{%
    \parbox{\linewidth}{%
        $<$s$>$[INST] $<<$SYS$>>$\\
        You are a teaching assistant. Answer questions from students on course logistics. First, choose the question type from one of these 7 options: 1) yes/no, 2) single factual, 3) multi factual, 4) single reasoning, 5) multi reasoning, 6) summarization, and 7) no answer. For questions of type single reasoning, provide a single reasoning step. For questions of type multi reasoning, provide up to 5 reasoning steps. For all other question types, don't provide any reasoning steps. Then, write the answer to the question. $<<$/SYS$>>$\\
        \{RAG context text for RAG baselines\}\\
        \#\#\# The question is: \{question text\}\\
        \#\#\# The question type is: [/INST]\\
    }%
}

\subsection{Prompts for GPT-4 with Retrieval Assistant}

For GPT-4 with Retrieval Assistant, we use the following system instructions:

\noindent\fbox{%
    \parbox{\linewidth}{%
        You are a teaching assistant. Answer questions from students on course logistics using the attached course syllabus document in your knowledge base. If an answer is not contained in the course syllabus output `No/insufficient information'.
    }%
}

Given the system instructions above, we use the following prompt for GPT-4 with Retrieval Assistant:

\noindent\fbox{%
    \parbox{\linewidth}{%
        The question is: \{question text\}\\
        The answer is:
    }%
}

\begin{table*}
    \centering
    \small
    \begin{tabular}{|p{1cm}|p{7cm}|p{7cm}|}
        \hline
        & \textbf{Fact-QA Precision} & \textbf{Fact-QA Recall}\\
        \hline
        \textbf{Prompt} & Your job is to evaluate the similarity of different answers to a single question. You will be given a question asking for information regarding a specific college course. You will also be given two possible answers to that question, and will have to evaluate the claims in one answer against the other.\newline

        Steps:\newline
        1. List all of the atomic claims made by Answer 1. Note that an answer saying that there is no information counts as a single claim.\newline
        2. Tell me which of those claims are supported by Answer 2.\newline
        3. Summarize the results using the template "Score: <num supported claims>/<num total claims>". Ensure that both numbers are integers.\newline
        
        Question: Can we work with our classmates on the homework assignments and if we do, will that affect our grade?\newline
        Answer 1: Yes you can, as long as you explictly mention the name of your collaborators for each problem. Also, you must write your own solution, otherwise it might affect your grade.\newline
        Answer 2: It is acceptable to work with others on the homework assignments but you must list them as having contributed to the work. The final answer has to be completed by you just like you would during an exam. You may receive a zero on the homework assignment if it appears that you did not solve the problem yourself. No programming code should be included in your answer. & Your job is to evaluate the similarity of different answers to a single question. You will be given a question asking for information regarding a specific college course. You will also be given two possible answers to that question, and will have to evaluate the claims in one answer against the other.\newline
        
        Steps:\newline
        1. List all of the atomic claims made by Answer 1. Note that an answer saying that there is no information counts as a single claim.\newline
        2. Tell me which of those claims are supported by Answer 2.\newline
        3. Summarize the results using the template "Score: <num supported claims>/<num total claims>". Ensure that both numbers are integers.\newline
        
        Question: Can we work with our classmates on the homework assignments and if we do, will that affect our grade?\newline
        Answer 1: It is acceptable to work with others on the homework assignments but you must list them as having contributed to the work. The final answer has to be completed by you just like you would during an exam. You may receive a zero on the homework assignment if it appears that you did not solve the problem yourself. No programming code should be included in your answer.\newline
        Answer 2: Yes you can, as long as you explictly mention the name of your collaborators for each problem. Also, you must write your own solution, otherwise it might affect your grade. \\
        \hline
        \textbf{Output} & 1. Atomic claims made by Answer 1:\newline
           - You can work with your classmates on the homework assignments.\newline
           - You must explicitly mention the name of your collaborators for each problem.\newline
           - You must write your own solution.\newline
           - Not writing your own solution might affect your grade.\newline
        
        2. Claims supported by Answer 2:\newline
           - You can work with your classmates on the homework assignments (supported by "It is acceptable to work with others on the homework assignments").\newline
           - You must explicitly mention the name of your collaborators for each problem (supported by "but you must list them as having contributed to the work").\newline
           - You must write your own solution (supported by "The final answer has to be completed by you just like you would during an exam").\newline
           - Not writing your own solution might affect your grade (supported by "You may receive a zero on the homework assignment if it appears that you did not solve the problem yourself").\newline
        
        3. Summarize the results:\newline
           Score: 4/4 & 1. Atomic claims made by Answer 1:\newline
           - It is acceptable to work with others on the homework assignments.\newline
           - You must list your classmates as contributors if they helped.\newline
           - The final answer must be completed individually.\newline
           - Not completing the final answer individually may result in receiving a zero.\newline
           - No programming code should be included in the answer.\newline
        
        2. Claims supported by Answer 2:\newline
           - It is acceptable to work with others on the homework assignments (implied by "Yes you can").\newline
           - You must list your classmates as contributors if they helped (implied by "as long as you explicitly mention the name of your collaborators").\newline
           - The final answer must be completed individually (implied by "you must write your own solution").\newline
           - Not completing the final answer individually may affect your grade (implied by "otherwise it might affect your grade").\newline
           - No programming code should be included in the answer (this claim is not addressed in Answer 2).\newline
        
        3. Score: 4/5\\
        \hline
    \end{tabular}
    \caption{Example prompts and outputs for Fact-QA Precision and F-Recall calculation for a single predicted answer. In each prompt, GPT-4 extracts the atomic facts from the first answer and compares each fact against the second answer, resulting in $1.0$ as the Fact-QA Precision and $0.8$ as the Fact-QA Recall. 
    }
    \label{tab:metric-prompt}
\end{table*}


\section{Dataset Collection: Annotator Instructions}
\label{sec:appendix_dataset_collection}

We provide the instructions given to annotators in Table~\ref{tab:instructions_qa}. We provide the tutorial given to annotators containing 14 diverse QA pairs based on an example syllabus from a course in Computer Science in Table~\ref{tab:tutorial_qa}.

\begin{table*}
\small
\centering
\scalebox{1}{
\begin{tabular}{p{0.95\linewidth}}

\toprule

        \textbf{Transforming education through AI: Write course logistics-related questions and answers}\\\\
        
        We need your help to transform education through AI. We’re a team of researchers building an AI-based chatbot to automatically answer students’ questions on course logistics. Instead of spending time answering mundane administrative questions, professors can focus their time and creativity on teaching.\\\\
        
        To train our AI-based chatbot, we need your help in writing logistics-related questions and answers on course syllabus. You’ll spend 60-90 minutes reading 1 course syllabus document and writing 14 questions and answers (QA). These 14 QAs will cover 7 types of questions with 2 QAs per type (14 = 7 X 2).\\\\

        The 7 question types are:\\\\
        1. Yes/no: These questions have a yes/no answer.\\
        2. Single-factual: The answers to these questions contain a single fact.\\
        3. Multi-factual: The answers to these questions contain multiple facts, possibly spread across the syllabus.\\
        4. Single-reasoning: These answers are not explicitly present in the syllabus and require a single reasoning step.\\
        5. Multi-reasoning: These answers are not explicitly present in the syllabus and require multiple reasoning steps.\\
        6. Summarization: These are open-ended answers requiring summarization of various information spread across the syllabus.\\
        7. No answer: These questions have no or insufficient information in the syllabus to answer them.\\

\bottomrule
\end{tabular}
}
\caption{Instructions provided to annotators asking them to simulate logistics-related QA pairs seen in real course classrooms.}
\label{tab:instructions_qa}
\end{table*}

\begin{table*}
\small
\centering
\scalebox{.73}{
\begin{tabular}{p{.3\linewidth}p{.7\linewidth}}

\toprule

\rowcolor{gray!21} \multicolumn{2}{c}{Yes/No}\\
\midrule
Question: & Is Introduction to Algorithms a required textbook in this course?\\
Answer span: & List of books or other equipment or technology  - No book required.\\
Answer: & No\\

\midrule

Question: & Can we use Java as the programming language for our assignments?\\
Answer span: & Programming languages to be used would include C/C++/Java and Python.\\
Answer: & Yes\\

\midrule
\rowcolor{gray!21} \multicolumn{2}{c}{Single-factual}\\
\midrule
Question: & What date is the midterm exam?\\
Answer span: & 11/14 Midterm (In-Class)\\
Answer: & The midterm exam will be on Nov 14.\\

\midrule

Question: & How many credits does this course provide?\\
Answer span: & credits: 3\\
Answer: & This course will provide 3 credits.\\

\midrule
\rowcolor{gray!21} \multicolumn{2}{c}{Multi-factual}\\
\midrule
Question: & When and where are TA office hours held?\\
Answer span 1: & Wed 10AM-11AM@CS207, Fri 9:30AM-10:30AM@CS207\\
Answer span 2: & Tu/Th. 4PM - 5PM@LGRT T220\\
Answer: & The following TA office hours will be held: 1) Every Wednesday from 10AM to 11AM at CS207. 2) Every Friday from 9:30AM to 10:30AM also at CS207. 3) Every Tuesday and Thursday from 4PM to 5PM at LGRT T220.\\

\midrule

Question: & What are the various sections to be included in our paper review?\\
Answer span 1: & For paper reviews, students will have to read a paper and submit a review having the following sections: Summary\\
Answer span 2: & Strengths of the system\\
Answer span 3: & Aspects that were not clear or hard to understand\\
Answer span 4: & Limitations of the system or possible extensions\\
Answer: & In your paper reviews, the following sections should be included: 1) Summary, 2) Strengths of the system, 3) Aspects that were not clear or hard to understand, and 4) Limitations of the system or possible extensions.\\

\midrule
\rowcolor{gray!21} \multicolumn{2}{c}{Single-hop reasoning}\\
\midrule
Question: & I’ve not taken COMPSCI 380. Do I satisfy the prerequisites?\\
Reasoning step: & COMPSCI 380 is not one of the prerequisites mentioned which are COMPSCI 377 operating systems and COMPSCI 445 information systems.\\
Answer: & Yes, you satisfy the prerequisites since the only prerequisites are COMPSCI 377 operating systems and COMPSCI 445 information systems.\\

\midrule

Question: & I will not be able to attend classes from Sep 28 to Oct 3. Which topics will I miss?\\
Reasoning step: & The topic of the class on Sep 28 is MapReduce and for the class on Oct 3 is Spark.\\
Answer: & You will miss topics on MapReduce and Spark.\\

\midrule
\rowcolor{gray!21} \multicolumn{2}{c}{Multi-hop reasoning}\\
\midrule
Question: & I’ll miss the midterm exam on Nov 14. Will I still be able to get an A-?\\
Reasoning step 1: & Midterm exam is $20\%$\\
Reasoning step 2: & Assuming perfect score in rest of the assessment components, the maximum score possible is $80/100$.\\
Reasoning step 3: & The minimum score for an A- is $85$.\\
Answer: & No, you will not be able to get an A- if you miss the midterm exam since A- requires a score of $85$ and above. The midterm exam accounts for $20\%$ of your grade. Assuming a perfect score in the rest of the assessment components (reviews, homework, quizzes and projects) the maximum score possible is $80$ if you miss the midterm exam.\\

\midrule

Question: & I really prefer to write paper reviews by hand. Will this impact my grade?\\
Reasoning step 1: & The instructions state ``please type; don’t handwrite; You are CS students. Any handwritten report will be treated as FAIL''.\\
Reasoning step 2: & Grading policy for paper review: fail = 50, pass = 100, missing or late = 0\\
Reasoning step 3: & Reviews account for $10\%$ of your grade.\\
Reasoning step 4: & Failing all reviews will affect $50\% \cdot 10\% = 5\%$ of your grade.\\
Answer: & Yes, this will impact your grade. You’ll lose $5\%$ of your grade if you write all reports by hand since they’ll all be treated as fails.\\

\midrule
\rowcolor{gray!21} \multicolumn{2}{c}{Summarization}\\
\midrule
Question: & Could you give me some details on exams in this course?\\
Answer span 1: & There will be a midterm exam. The midterm exam will only be about the material discussed before the exam.\\
Answer span 2: & Midterm exam is in-class.\\
Answer span 3: & No final exam. Instead, there will be project presentations at the end of the semester.\\
Answer: & There will only be a midterm exam in-class based on the material before the exam date. In lieu of the final exam there will be a project presentation at the end of the semester.\\

\midrule

Question: &  What are the administrative details on quizzes?\\
Answer span 1: & There will be weekly quizzes about the material discussed in the week.\\
Answer span 2: & The quiz will be on Monday.\\
Answer span 3: & Submission requirements: On Moodle\\
Answer span 4: & Grading policy: Each quiz has the same weight. The averaged score will be used as the final score.\\
Answer: & Quizzes will be conducted every Monday on Moodle based on the material discussed in the week. Each quiz has the same weight and the average of all quizzes will be used for grading.\\

\midrule
\rowcolor{gray!21} \multicolumn{2}{c}{Adversarial}\\
\midrule
Question: & Will course lectures be recorded and published online?\\
Answer: & No/insufficient information\\

\midrule

Question: & Is there any penalty for missing class lectures?\\
Answer: & No/insufficient information\\

\bottomrule
\end{tabular}
}
\caption{Tutorial provided to annotators containing a diverse set of 14 QA pairs across all 7 question types, along with a reference syllabus from Computer Science.}
\label{tab:tutorial_qa}
\end{table*}


\section{Misc: Attribution}
Icons used in Figure~\ref{fig:diverse_majors} are from \url{www.flaticon.com}.

\end{document}